\documentclass[11pt]{article}
\pdfoutput=1

 \usepackage{jcappub}

\usepackage[table]{xcolor}
\usepackage{makecell}
\usepackage{caption}
\usepackage{subcaption}
\usepackage{hyperref}
\usepackage{graphicx}
\usepackage{amsmath}
\usepackage{mathtools}
\usepackage{physics}
\usepackage{siunitx}
\usepackage{tablefootnote}
\usepackage{soul}
\usepackage{color}
\usepackage{verbatim}
\usepackage{float}
\usepackage{bm}
\usepackage{ulem}
\usepackage{orcidlink}
\usepackage[nameinlink, capitalise]{cleveref}
\usepackage{braket}
\usepackage{dirtytalk}
\usepackage{mathtools}
\usepackage{amsmath}
\usepackage{tikz} 
\usepackage{tikz-feynman}
\tikzfeynmanset{compat=1.1.0}
\usepackage{feynmp}
\usetikzlibrary{shapes.geometric, backgrounds, arrows, calc, fit}
\tikzstyle{process} = [rectangle, rounded corners, minimum width=8cm, minimum height=1cm, text centered, draw=black, fill=blue!30]
\tikzstyle{arrow} = [thick,->,>=stealth]
\usepackage{underscore}

\usepackage{color,appendix,latexsym,amsmath,amssymb,graphicx,booktabs,epsfig,hyperref,url}

\numberwithin{equation}{section}
\usepackage[english]{babel}

\usepackage{letltxmacro}
\LetLtxMacro{\originaleqref}{\eqref}
\usepackage[sorting=none]{biblatex} 
\addto\extrasenglish{%
}

\definecolor{MyBlue}{rgb}{0.15,0.15,0.70}
\hypersetup{
colorlinks=true,
citecolor=MyBlue,
linkcolor=MyBlue,
urlcolor=MyBlue
}

\definecolor{orange}{rgb}{0.98, 0.6, 0.01}
\definecolor{darkolivegreen}{rgb}{0.33, 0.42, 0.18}
\definecolor{tealblue}{rgb}{0.21, 0.46, 0.53}

\usepackage{xspace}

\newcommand{\fpbh}{\ensuremath{f_{\rm PBH}}}

\usepackage{listings}
\definecolor{codegreen}{rgb}{0,0.6,0}
\definecolor{codegray}{rgb}{0.5,0.5,0.5}
\definecolor{codepurple}{rgb}{0.58,0,0.82}
\definecolor{backcolour}{rgb}{0.95,0.95,0.92}
\definecolor{darkgreen}{rgb}{0.0, 0.6, 0.0}

\title{Comprehensively Constraining Ultra--Light Primordial Black Holes Through Relic Formation and Early Mergers}

\author[1,2]{Amirah Aljazaeri \,\orcidlink{0009-0008-7032-8286},}
\author[1]{Christian T.~Byrnes\,\orcidlink{0000-0003-2583-6536},}

\affiliation[1]{Department of Physics and Astronomy, University of Sussex, Brighton BN1 9QH, UK\\}
\affiliation[2]{
Department of Physics, Taibah University, Madinah 42353, Saudi Arabia }

\emailAdd{aa2409@sussex.ac.uk}
\emailAdd{C.Byrnes@sussex.ac.uk}

\date{}

\abstract{We investigate constraints on the least explored, smallest mass scales of primordial black holes (PBHs), which evaporate prior to Big Bang Nucleosynthesis (BBN). Our study examines the impact of Planck-mass relics on the allowed fraction of dark matter composed of PBHs ($f_{\mathrm{PBH}}$), as well as on the resulting stochastic gravitational wave background and the formation of primordial binaries. We discuss how these binaries and early mergers lead to longer PBH lifetimes, extending the reach of the stringent BBN constraints to smaller masses than usually expected.  We make comprehensive constraint plots on the collapse fraction $\beta$ and $f_{\mathrm{PBH}}$ (including relics) focusing on ultra-light PBHs.
}

\begin{document}

\maketitle
\flushbottom

\section{Introduction}
\label{secintroduction}
Primordial black holes (PBHs), initially proposed by Zel'dovich et al. \cite{zel1966hypothesis} and further developed by Hawking and collaborators \cite{hawking1971gravitationally,carr1974black}, 
are theoretical black holes that could have formed in the early universe through the collapse of large density perturbations which may arise in inflationary models that produce pronounced peaks in the primordial power spectrum. For detailed reviews, see \cite{carr2021constraints, sasaki2018primordial}. The role of PBHs has gained interest following suggestions that some of the LIGO-Virgo detections of GWs may originate from binary PBHs \cite{sasaki2016primordial,bird2016did,clesse2017clustering
}, and their potential to explain dark matter without invoking new particles nor new physics \cite{chapline1975cosmological}.
 
 During the radiation-dominated era, PBHs could form within a broad range of masses, with their initial mass proportional to the horizon mass at the time of horizon entry, 
 \begin{equation}
    \label{eqMPBHi}
    M_{\rm{PBH,i}} = \frac{4 \pi}{3} H^{-3} \epsilon_{\rm{r}}  \simeq  1 \times 10^{39} 
    \left(\frac{t_{\rm{i}}}{\rm{s}}\right)
    \,\,\rm{g},
\end{equation}
where $H^{-1}$ corresponds to the Hubble radius at the time of formation, and $\epsilon_{\rm{r}}$ is the radiation energy density. Throughout this paper, for simplicity, we approximate the PBH mass distribution as nearly monochromatic and neglect the impact of accretion, as its effects are likely negligible on the small scales relevant to this work. The lifetime of a PBH, $ t_{\rm{ev}} $, is crucial for its evolution. For PBHs evaporating before Big Bang Nucleosynthesis (BBN), assuming a constant relativistic degrees of freedom $g_{*,H}(T_{BH}) \approx 108$ \cite{hooper2019dark}, the lifetime is:
\begin{equation}
\label{eqtev}
   t_{\rm{ev}} \simeq 4 \times 10^{-28}
    \left(\frac{M_{\rm{PBH,i}}}{\rm{g}}\right)^{3} \,\,\,\rm{s}.
\end{equation}
Therefore, PBHs with initial mass $M_{\rm{PBH,i}} \gtrsim 10^{15}$g have lifetimes exceeding the current age of the universe. 

Given the broad range of possible initial PBH masses, various constraints have been established to evaluate their impact on nearby astronomical objects (for a historical overview, see \cite{chapline1975cosmological, hawking1975particle, carr2010new}), including phenomena such as microlensing and the energy injection associated with the evaporation of lighter PBHs (for a review of the constraints on evaporating PBHs, see \cite{auffinger2023primordial}). These constraints indicate that PBHs with a monochromatic mass distribution cannot account for the entirety of dark matter, except within a specific mass range, approximately $ 10^{17} \rm{g} \lesssim M_{\rm{PBH}} \lesssim 10^{22} \rm{g} $, (see \cite{carr2021constraints}, and references therein). Nevertheless, efforts have been made to bypass these constraints.
For instance, corrections may arise when considering Hawking's semi-classical approximation, called ``memory burden"\cite{dvali2020black}. It is argued that Hawking evaporation becomes ineffective when the black hole mass drops to about half its initial value, potentially opening a new window for considering light PBHs as dark matter candidates; see e.g~\cite{thoss2024breakdown,Montefalcone:2025akm,Dvali:2025ktz}. 

Alternatively, after evaporating most of their mass, PBHs may leave behind stable Planck-mass relics \cite{Carr1994,Chen2005,Green1997} at the intersection of quantum mechanics, relativity, and Newtonian gravity, potentially invalidating standard physics for lower-mass PBHs \cite{garfinkle2009planck}, allowing them to constitute all of the dark matter of today. Although several arguments support the existence of such stable remnants \cite{lehmann2019direct, domenech2023gravitational}, complete evaporation remains a viable possibility due to unknown Planck-scale physics, which fundamentally makes the issue a question of quantum gravity \cite{kazemian2023diffuse}.
Detecting those PBH Planckian relics is challenging unless they carry an electric or magnetic charge \cite{lehmann2019direct,profumo2024ultralightprimordialblackholes} or undergo substantial merging and evaporation in the late universe. A promising alternative observable is stochastic gravitational waves background  (SGWB), which can provide indirect insights into the formation and abundance of both ultra-light PBHs. 

In this paper, we examine current cosmological and astrophysical constraints on PBHs, including the possibility of Planck-mass relics. We consider a potential early PBH-dominated (ePBHd) epoch and the generation of gravitational waves (GWs) from second-order cosmological perturbations theory, analyzing SGWB sources and their constraints on PBH abundance.  Additionally, we explore the implications of primordial PBH binary formation and mergers, including their impact on the GW energy density. Finally, we investigate the consequences of multiple successive mergers occurring just before BBN and derive new constraints on the initial PBH mass spectrum to ensure consistency with BBN bounds.

This paper is structured as follows: In \cref{secbetaConstraints}, we define the abundance of PBHs and introduce the concepts of Planckian relics and the epoch of ePBHd. We then explore the resulting constraints, including those from SGWs (\cref{secGWs}) and primordial binary mergers (\cref{secMerger}), which are collectively summarized in a comprehensive plot in \cref{secbetaplot}. Building on this foundation, we analyze the constraints on the fraction of dark matter composed of PBHs (\cref{secfPBH}), highlighting how the assumption of Planckian relics strengthens the constraints (\cref{sectighterCons}), and how the ePBHd further constrains the PBH dark matter fraction (\cref{secmappingONf}). Finally, we explore the less well-known parameter window in which all dark matter could consist of Planckian relics from evaporated PBHs (\cref{secrelicsWindow}). Throughout our analysis, we adopt the (non-reduced) Planck mass, $M_{\rm{pl}} = \sqrt{\hbar c/G} \approx 2 \times 10^{-5} \, \mathrm{g}$, as the benchmark scale for relic formation.

\section{Constraining the abundance of Primordial Black Holes}
\label{secbetaConstraints}
Although the existence of evaporating PBHs today is tightly constrained, the study of ultra-light PBHs offers significant advantages by potentially unveiling new avenues for understanding the origins of dark matter, either through Hawking radiation \cite{lennon2018black,morrison2019melanopogenesis,hooper2019dark} or whether evaporation leads to the formation of a stable relic \cite{Carr1994,Chen2005,Green1997}, or by enhancing our understanding of the dynamics of the early universe, including a period of ePBHd \cite{hooper2019dark,anantua2009grand}, baryogenesis \cite{fujita2014baryon}, and the production of dark radiation \cite{hooper2019dark,arbey2021precision}.

The parameter $\beta_i$ represents the fraction of the universe collapsing into PBHs at formation and is typically estimated assuming instantaneous formation after horizon re-entry. We assume PBHs form during the radiation-dominated era, where their relative density compared to the background scales with the scale factor ($a$) until matter-radiation equality.  More generally, $\beta$ is defined as $\rho_{\rm{PBH}}/\rho_{\rm{tot}}$ at any given time. Focusing on PBH formation, and assuming no change in the PBH mass and no early matter dominated era (for now), we express it as:
\begin{align}
    \beta_{\rm{i}} &= \frac {\rho_{\rm{PBH}}}{\rho_{\rm{tot}}} \, \Bigg|_{\rm{form}} 
    = \frac {\rho_{\rm{PBH}}^{\rm{i}}}{\rho_{r}} \,
    = \frac{\rho_{\rm{PBH}}^{\rm{eq}}}{\rho_{\rm{tot}}^{\rm{eq}}} \left(\frac{a_{\rm{i}}}{a_{\rm{eq}}}\right)\,
    \approx \frac{f_{\rm{PBH}}}{2\left(1+ \frac{\Omega_{\rm{b}}}{\Omega_{\rm{DM}}}\right)} \left(\frac{a_{\rm{i}}}{a_{\rm{eq}}}\right) \notag \\
    & \approx 1.13 \times 10^{-26} \left(\frac{M_{\rm{PBH,i}}}{\rm{g}}\right)^{1/2} f_{\rm{PBH}},
    \label{eqBetaNonevap}
\end{align}
where the factor $2(1+\Omega_{\rm{b}}/\Omega_{\rm{DM}})$ accounts for both dark matter and baryon contributions to the total matter density at equality, and $f_{\rm{PBH}}$ is the present fraction of PBHs in dark matter. For simplicity, we neglect changes in the number of degrees of freedom; see \cite{carr2010new,green2021primordial} for more detailed expressions of $\beta$; however,  this approximation is sufficient for our purposes. We also use the horizon mass at matter-radiation equality, 
\begin{align}
\label{eqMHeq}
    M_{\rm{H}}^{\rm eq} 
    &= \frac{4 \pi}{3} \rho_{\rm{eq}} H_{\rm{eq}}^{-3} c^{3} \notag \\
    & \simeq 2.9 \times 10^{17} \, \rm{M}_{\odot},
\end{align}  
where $\rho_{\rm{eq}}= 2 \Omega_{\rm{r}} \rho_{\rm{c}} a_{\rm{eq}}^{-4}$ and $H_{\rm{eq}} = c \, k_{\rm{eq}} / a_{\rm{eq}}$, with $k_{\rm{eq}} \sim 0.01 \, \rm{Mpc}^{-1}$, $\Omega_{\rm{r}} \sim 9 \times 10^{-5}$, and $H_0 \sim 2.2 \times 10^{-18} \, \rm{s}^{-1}$ \cite{ade2016planck}.

\begin{figure}[ht]
    \centering
    \begin{minipage}{0.56\textwidth} 
        \centering
        \includegraphics[width=1\linewidth]{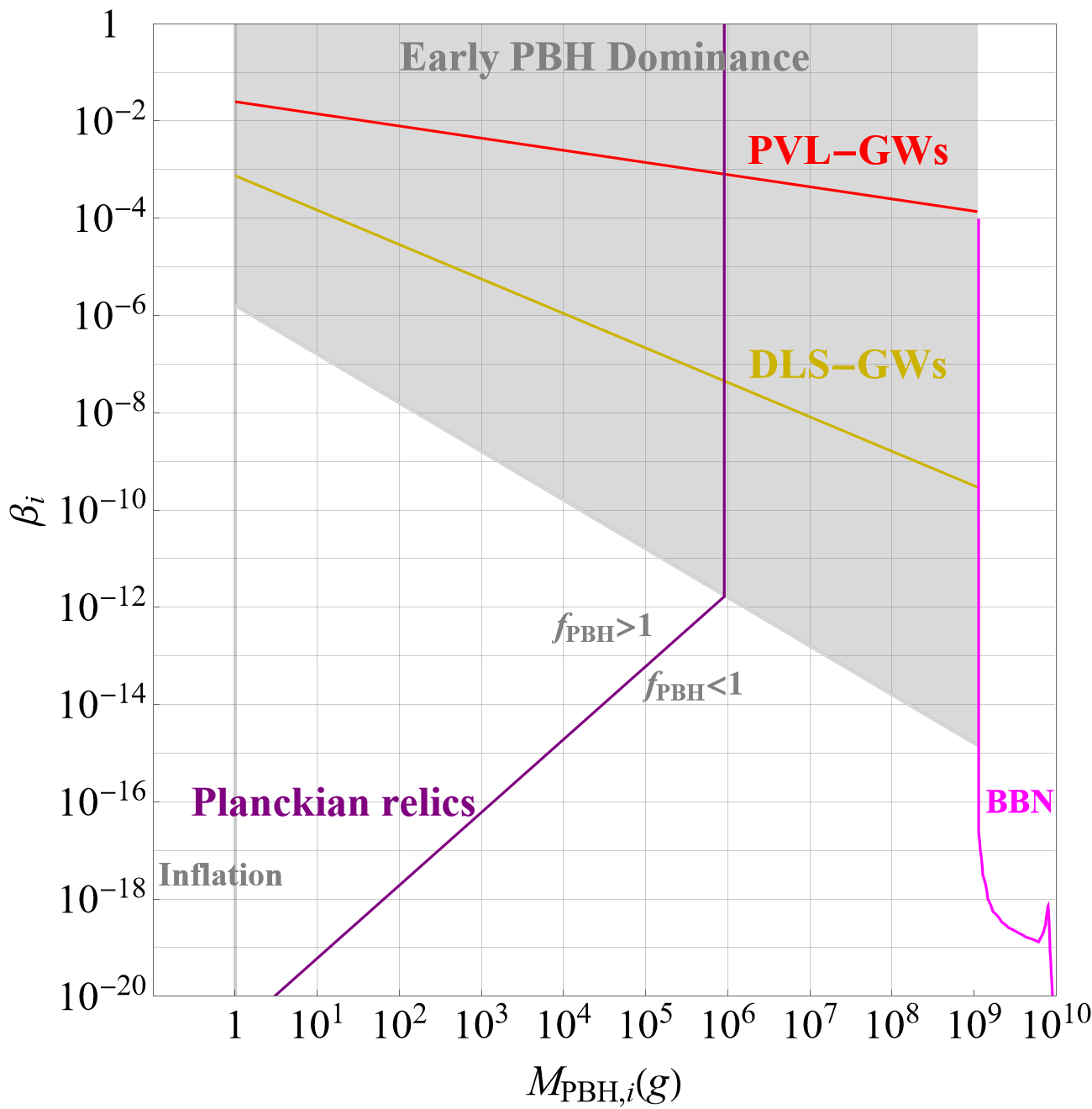}
    \end{minipage}%
    \hspace{1em} 
    \begin{minipage}{0.4\textwidth} 
        \caption[...]{Constraints on the initial abundance, $\beta_{\rm{i}}$, of ultra-light PBHs as a function of the initial mass, $M_{\rm{PBH,i}}$, from stochastic gravitational waves background (\cite{papanikolaou2021gravitational} in red, \cite{domenech2021gravitational} in yellow), evaporation into Planck-mass relics assuming $f_{\rm PBH} \leq 1$ (purple), and BBN constraints (\cite{carr2021constraints}, magenta). The grey-shaded region indicates early PBH-domination, and the grey line at $M_{\rm{PBH,i}} = 1$g marks the minimum mass of PBHs formed post-inflation.}
        \label{figGWs}
    \end{minipage}
\end{figure}

To account for Planck-mass relics left behind by evaporating PBHs, we need to include the ratio of the final relic mass to the initial PBH mass in \cref{eqBetaNonevap}, leading to:
\begin{equation}
    \label{eqbetaD}
    \beta_{\rm{relic, i}} \approx 5.65 \times 10^{-22} \left(\frac{M_{\rm{PBH,i}}}{{\rm{g}}}\right)^{3/2} f_{\rm{PBH}}, 
\end{equation}  
which, based on the constraint $ f_{\rm PBH} \leq 1 $  and the assumption that such relics could exist, is represented by the purple line in \cref{figGWs}. 

In general, the different scaling of matter (PBHs) and radiation with the scale factor $a$ can increase the PBH abundance ($\beta$) by up to 25 orders of magnitude, depending on their initial mass. Thus, even a small initial fraction may grow enough to trigger an ePBHd epoch. If this epoch starts at time $t_{\rm{d}}$ and scale factor $a_{\rm{d}}$, PBHs dominate when $a_{\rm{d}} = a_{\rm{i}} / \beta_{\rm{i}}$, since their energy density fraction ($\beta = \rho_{\rm{PBH}} / \rho_{\rm{tot}}$) grows linearly with $a$. Therefore, the time when PBHs first dominate is
\begin{equation}  
    \label{eqtd}  
    t_{\rm{d}} = \frac{t_{\rm{i}}}{\beta_{\rm{i}}^2}.  
\end{equation}
During this phase, PBHs will dominate over all components, which, by definition, means $\beta \simeq 1$. The condition for this to occur is 
\begin{equation}
    \label{eqBibiggerBcri}
     \beta_{\rm{i}} > \beta_{\rm{cri}} = \left(\frac{t_{\rm{i}}}{t_{\rm{ev}}}\right)^{1/2} 
     \simeq 1.6\times 10^{-6} \left(\frac{M_{\rm{PBH,i}}}{\rm{g}}\right)^{-1}.
\end{equation}
 
It should be noted that this critical value represents a limit (see the shaded gray area in \cref{figGWs}), not an observational constraint on $\beta_{\rm{i}}$. It enhances the amplitude of SGWs and constrains $f_{\rm{PBH}}$, as discussed later. Significant efforts have focused on constraining the parameter $\beta_i$, particularly for non-evaporated PBHs, with the tightest bounds highlighted in \cite{green2015primordial} and also in \cite{agius2024feedback,pritchard2025constraining}. While \cite{carr2021constraints} provides careful indirect constraints, stronger constraints are expected, especially from SGWs during the ePBHd period (see \cref{secGWs}). Our goal in this section is to present a comprehensive plot that integrates the strongest indirect constraints (see \cref{secbetaplot}).

\subsection{Stochastic gravitational wave background constraints on the abundance of PBH}
\label{secGWs}
Hawking evaporation must complete before BBN to remain consistent with observational constraints, unless $\beta_i \ll 1$. If a PBH fully evaporates, it leaves no direct signatures unless a Planck-mass relic remains. The absence of observed Hawking evaporation places constraints on certain PBH scenarios. Evaporation can also produce dark radiation, such as axions, which contributes to the total radiation density, typically parameterized by the effective number of neutrino species, $N_{\rm{eff}}$, which can be modified in an ePBHd era \cite{hooper2019dark, hooper2020hot}. Gravitational waves are also a valuable probe for studying early cosmic phenomena and constraining the evolution of the universe. 

The potential relevance of PBHs in the study of SGWs has attracted significant attention, with contributions arising from various sources. During cosmic inflation, overdense perturbations that lead to PBH formation can induce second-order GWs, as described by the cosmological perturbation theory \cite{acquaviva2002second, nakamura2019second}. Additionally, random fluctuations in the spatial distribution of PBHs contribute further to the SGWB \cite{papanikolaou2021gravitational, domenech2021gravitational}. SGWs can also originate from PBH-related processes such as Hawking radiation, including graviton emission during PBH evaporation \cite{dolgov2011relic, ireland2023primordial}, as well as from PBH mergers \cite{raidal2017gravitational, zagorac2019gut, pujolas2021prospects}. However, the most significant impact of PBH mergers is primarily associated with solar-mass PBHs, while the contribution from ultra-light PBHs is discussed later (see \cref{secMerger}). In this section, we examine these various sources of SGWs in more detail.

Papanikolaou, Vennin and Langlois (PVL) in 2021 made the first constraint on the abundance of very light PBHs \cite{papanikolaou2021gravitational} which is independent of whether PBHs leave relics. Assuming that PBHs  are distributed randomly, following Poisson statistics, they showed that the associated growth of the power spectrum and isocurvature fluctuations leads to the generation of SGWs afters PBHs dominate with peak amplitude
\[
\Omega_{\rm{GW}} \simeq 10^{10} \left(\frac{M_{\rm{PBH,i}}}{{\rm{g}}}\right)^{\frac{4}{3}} \beta_{\rm{i}}^{\frac{16}{3}}.
\]
Their resulting constraint was 
\begin{equation}
\label{eqFranceGWs}
    \beta_{\rm{i}} \lesssim 1.4 \times 10^{-4}  \left(\frac{M_{\rm{PBH,i}}}{{10^{9}\,\rm{g}}}\right)^{-1/4},
\end{equation}
which was derived by analyzing the GW backreaction problem.

However, the validity of this constraint is uncertain, as density perturbations will eventually become nonlinear. To account for this, following \cite{PhysRevD.105.123023}, we incorporate a suppression factor $ R $ which depends on both $M_{\rm{PBH,i}}$ and $\beta_{\rm{i}}$ and satisfies  
\begin{equation}
\label{eqR}
R \le  \left( \left(\frac{t_{\rm{ev}}}{t_{\rm{d}}}\right)^{-2/3} 10^{-2} \sqrt{\frac{3 \pi}{2}}\frac{1}{\beta_{\rm{i}}}  
\right)^{2} \simeq 2.4 \times 10^{-18} \left(\frac{M_{\rm{PBH,i}}}{\rm{g}}\right)^{-8/3} \beta_{\rm{i}}^{-14/3}.
\end{equation} 
Applying this suppression to the peak GW amplitude, we analytically found that the expression for $\Omega_{\rm{GW}} \times R$ is a decreasing function of mass. Using this, we determine the maximum possible GW amplitude by assuming the maximum possible abundance ($\beta_{\rm{i}} = 1$) and the smallest possible mass ($M_{\rm{PBH,i}} = 1 \text{ g}$), yielding  
\[
\Omega _{\rm{GW}}^{\rm{sup}} \simeq 10^{-12} \left(\frac{M_{\rm{PBH,i}}}{{\rm{g}}}\right)^{-\frac{4}{3}} \beta_{\rm{i}}^{\frac{2}{3}} \simeq 2.4\times 10^{-12}.
\]  
Thus, regardless of the PBH mass, the resulting GW signal becomes negligible. However, as stated in \cite{PhysRevD.105.123023}, cutting off GW generation when the density perturbation grows non--linear is almost certainly too conservative and this issue deserves further study.

Another source of SGWs, proposed by Dom{\`e}nech, Lin, and Sasaki (DLS)  \cite{domenech2021gravitational}, suggests that induced GWs during the ePBHd period are generated immediately after evaporation, when the curvature perturbation undergoes dramatic fluctuations due to the sudden transition to a radiation-dominated universe \cite{inomata2017inflationary}. This abrupt transition leads to a significantly enhanced GW spectrum, imposing a stricter constraint on PBH abundance than \cref{eqFranceGWs}, given by:  
\begin{equation}
\label{eqSasakiGWs}
    \beta_{\rm{i}} \lesssim 1.1 \times 10^{-6}  \left(\frac{M_{\rm{PBH,i}}}{{10^{4}\,\rm{g}}}\right)^{-17/24}.
\end{equation}  
This constraint is highly sensitive to the PBH mass function (A monochromatic mass function appears to be assumed in their approach), which is implicitly suggested to fall within a very narrow log-normal width of $\lesssim 0.01$ as otherwise the transition becomes markedly less rapid \cite{inomata2020gravitational} - see also \cite{pearce2024gravitational} which shows the GW signal rapidly decreases for slower transitions. 

Even within a monochromatic power spectrum, such a narrow mass function is unrealistic, as $M_{\rm{PBH,i}}$ depends on the overdensity, following $M_{\rm{PBH,i}} \approx M_{\rm{H}}(\delta - \delta_c )^{\gamma}$, where $\gamma$ is the critical collapse exponent. If the density contrast slightly exceeds the critical threshold, PBH formation is delayed, yielding smaller black holes as outer layers escape. Thus, even for a fixed generating length scale ($M_{\rm{H}}$), $M_{\rm{PBH,i}}$ exhibits some variation, introducing a finite mass function width \cite{Escriva:2022duf}. Consequently, the constraint of \cite{domenech2021gravitational} may not reflect a realistic physical scenario, and dissipative effects will also weaken the signal \cite{Domenech:2025bvr}. Moreover, as we will show, early PBH mergers may weaken this constraint. Nonetheless, \cite{domenech2021gravitational} identifies an initial PBH mass range that could produce a detectable GW signal if the mass function remains sufficiently narrow. Finally, these constraints on $\beta_{\rm{i}}$ derived in \cref{eqFranceGWs} and \cref{eqSasakiGWs} are shown in \cref{figGWs} as a function of the initial PBH mass.

As a final takeaway of this subsection, \cref{tabGWsources} summarizes the main GW sources associated with PBHs, along with their corresponding peak frequencies and present day amplitudes. The relation between the frequency observed today and the comoving wavenumber is given by $f_0 = c\,k/2\pi a_0$. Moreover, while gravitational waves can also be generated by the collapse of large primordial curvature perturbations \cite{Domenech2021}, this source is not included in the table, as it does not rely on the presence of PBH and strongly depends on the specific formation mechanism.

\begin{table}[h!]
    \centering
    \makebox[\textwidth]{ 
        \resizebox{1\textwidth}{!}{ 
            \begin{tabular}{|p{2.5cm}|p{4.5cm}|p{7cm}|p{5cm}|}
                \hline
                \makecell{\textbf{GWs source}} & \makecell{\textbf{Explanation}} & \makecell{\textbf{peak $f$ (Hz) today}} & \makecell{\textbf{peak amplitude today}} \\ \hline
                \rowcolor{gray!20} PBH density 
                
                perturbations & The randomness comes from rare PBH formation \cite{papanikolaou2021gravitational} & \makecell{\\$f \simeq 10^{8} \beta_{\rm{i}}^{\frac{2}{3}} \left(\frac{M_{\rm{PBH,i}}}{{\rm{g}}}\right)^{-\frac{5}{6}}$ to $ 10^{7} \left(\frac{M_{\rm{PBH,i}}}{{\rm{g}}}\right)^{-\frac{5}{6}}$ \\
                provided $\beta_{\rm{i}} \ll 1$} & \makecell{$\Omega_{\rm{GW}} \simeq 10^{10} \left(\frac{M_{\rm{PBH,i}}}{{\rm{g}}}\right)^{\frac{4}{3}} \beta_{\rm{i}}^{\frac{16}{3}} $} \\ \hline
                \rowcolor{gray!20} PBH density 
                
                perturbations -suppressed & Suppressed by accounting for the impact of going nonlinear in density perturbations \cite{PhysRevD.105.123023} & \makecell{The GW amplitude is negligible} & \makecell{\\$\Omega_{\rm{GW}}^{\rm{sup}}  \simeq 10^{-12} \left(\frac{M_{\rm{PBH,i}}}{{\rm{g}}}\right)^{\frac{-4}{3}} \beta_{\rm{i}}^{\frac{2}{3}}$} \\ \hline
                \rowcolor{gray!20} Rapid 
                
                transition & From matter domination to radiation domination \cite{domenech2021gravitational} & \makecell{\\ $f \simeq 10^{6} \left(\frac{M_{\rm{PBH,i}}}{{\rm{g}}}\right)^{-\frac{5}{6}}$} & \makecell{\\$\Omega_{\rm{GW}} \simeq 10^{20} \left(\frac{M_{\rm{PBH,i}}}{{\rm{g}}}\right)^{\frac{14}{3}} \beta_{\rm{i}}^{\frac{16}{3}}$} \\ \hline
                Gravitons & The Hawking evaporation of PBHs into gravitons \cite{dolgov2011relic} & \makecell{$f \simeq 10^{13} \left(\frac{M_{\rm{PBH,i}}}{{\rm{g}}}\right)^{\frac{1}{2}}$} & \makecell{$\Omega_{\rm{GW}}\simeq 10^{-7} $} \\ \hline
                \rowcolor{gray!20} PBH merger & PBHs move and capture each other randomly in the early universe \cite{hooper2020hot} & \makecell{\\$f \simeq 10^{15} \left(\frac{M_{\rm{PBH,i}}}{{\rm{g}}}\right)^{\frac{1}{2}}$} & \makecell{\\$\Omega_{\rm{GW}}\simeq 10^{-5}$} \\ \hline
            \end{tabular}
        }
    }
    \caption{Summary of relevant induced GW sources for PBHs which evaporate before BBN. The equations are simplified by adopting representative values for the effective d.o.f.: $g_{\rm{eff}} = 100$, $g_* = 106.75$, the spin-weighted d.o.f. up to the PBH temperature $g_{\rm{H}}(T_{\rm{PBH}}) = 108$, and the entropic effective d.o.f. $g_{*,\rm{s}} = 106.75$. We follow the d.o.f. notation used in the original references. Sources requiring an early PBH-dominated era are shaded in gray. 
    }
    \label{tabGWsources}
\end{table}

\subsection{Primordial binary formation and merger}
\label{secMerger}
If the PBH evaporation timescale, $t_{\rm{ev}}$, exceeds the PBH merger timescale, $t_{\rm{mer}}$, then the impact of mergers must be considered. Specifically, when the initial PBH fraction, $\beta_{\rm{i}}$, is large enough to induce an ePBHd era, many PBHs are expected to form binaries and merge before evaporating. This changes observational constraints in a way not previously taken into account, potentially weakening those based on the rapid transition from matter to radiation domination by making the mass function less monochromatic but strengthening constraints for PBHs whose lifetime is sufficiently extended by the merger(s) to interfere with BBN. 

Many studies of PBH mergers, including \cite{nakamura1997gravitational, vaskonen2020lower, ding2023merger}, focus on a specific subset of mergers which merge today. In contrast, we are interested in understanding when typical binaries form after PBHs formation. To approach this, we need to understand when binary formation takes a place; see \cref{figBinary}. A binary system containing, for simplicity, two equal mass black holes ($M_{\rm{BH}} \equiv M_{\rm{PBH,i}}^{\rm{1}} = M_{\rm{PBH,i}}^{\rm{2}}$) may form at time, $t_{\rm{bf}}$, when the Newtonian gravitational attraction force between them balances the acceleration of the background expansion during radiation domination:
\begin{figure}
    \centering
    \includegraphics[width=12cm]{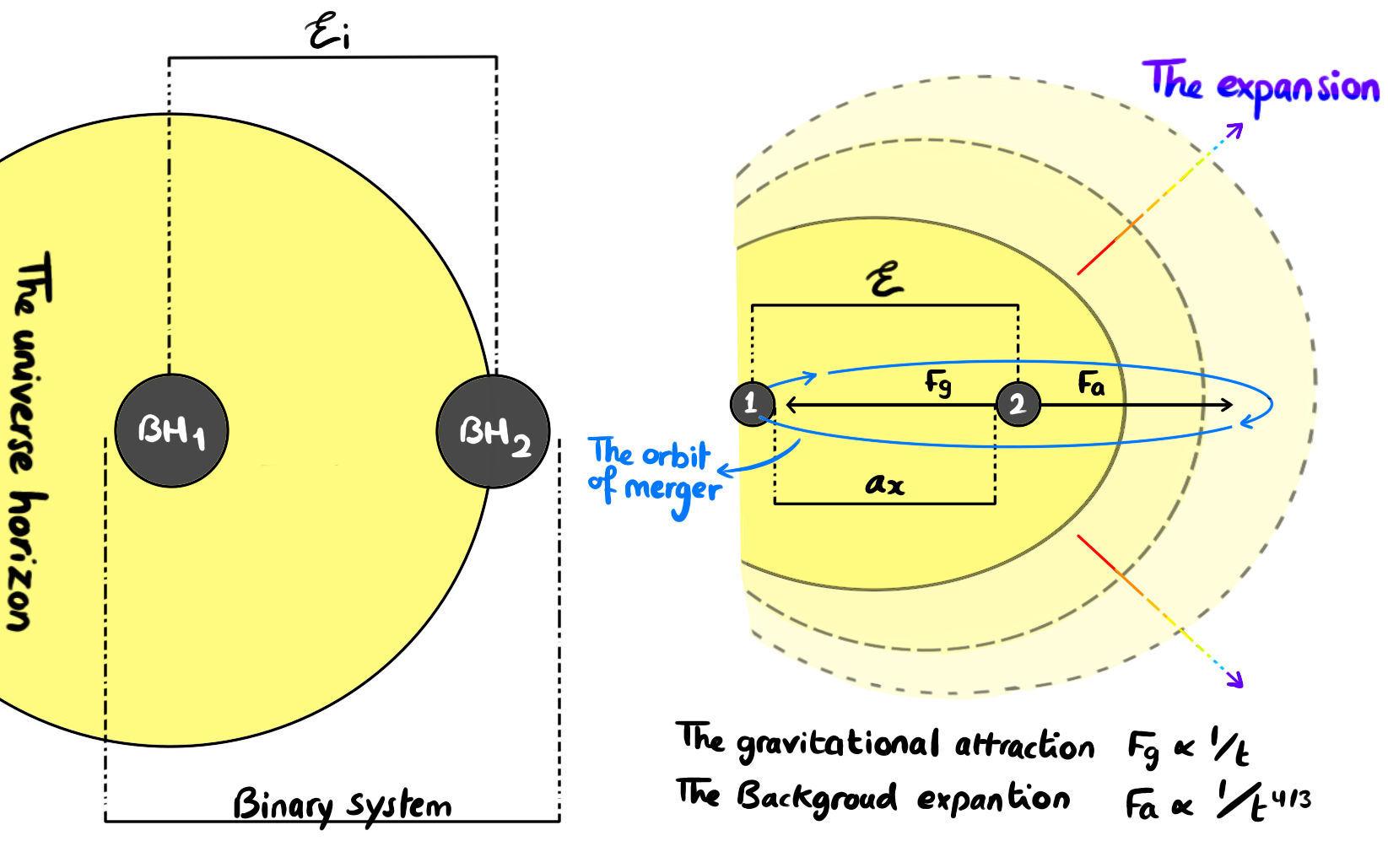}
    \caption[...]{\label{figBinary} Primordial binary formation. Once a binary system of two PBHs with an initial typical separation ($\bar{\xi}_{\rm{i}}$) enter the horizon, they (optimistically) can form a binary. However, a more realistic scenario is that the separation between the PBHs grows  due to cosmic expansion until the gravitational force ($F_{\rm{g}}\propto1/t$) between them balances the acceleration of the background expansion ($F_{\rm{a}}\propto 1/t^{4/3}$) during radiation domination.
    }
\end{figure}
\begin{equation}\label{eqbalance}
    \frac{G M_{\rm{PBH,i}}}{\bar{\xi}^{2}} = \bar{\xi} H^{2}
    ,
\end{equation} 
where $H = 1/2t$, $G$ is the gravitational constant, the typical separation as a function of its typical initial separation is ($\bar{\xi} = \bar{\xi}_{\rm{i}}\,  a_{\rm{bf}}/a_{\rm{i}}$), and the initial separation is:
\begin{equation*}
    \bar{\xi}_{\rm{i}} = \left(\frac{3 M_{\rm{PBH,i}}}{4 \pi \rho_{\rm{R}}^{\rm{i}} \beta_{\rm{i}}}\right)^{1/3},
\end{equation*}
where we used, for a sphere surrounding a black hole of mass,  $M_{\rm{PBH,i}}$, the total mass, $M_{\rm{tot}} = M_{\rm{PBH,i}} + \frac{4\pi}{3}\rho_{\rm{R}}^{\rm{i}}\bar{\xi}_{\rm{i}}^{3}$, $\rho_{\rm{tot}} = \rho_{\rm{PBH}}^{\rm{i}} + \rho_{\rm{R}}^{\rm{i}} = \rho_{\rm{R}}^{\rm{i}}(\beta_{\rm{i}}+1)$, and the overhead bar denotes the average value \cite{ding2023merger}. 
Therefore, substituting this into the force balance equation, the average time for binary to form will be
\begin{equation}
    \label{eqtbf}
    t_{\rm{bf}} = \left(\frac{\bar{\xi}_{\rm{i}}^{3}}{4 G M_{\rm{PBH,i}} t_{\rm{i}}^{3/2}}\right)^{2}\simeq 4 \frac{t_{\rm{i}}}{\beta_{\rm{i}}^2},
\end{equation}
which occurs near the time when PBHs begin to dominate the universe, $t_{\rm{d}} = t_{\rm{i}}/\beta^2_{\rm{i}}$. Consequently, binary formation typically takes place at the latest possible moment, around $t_{\rm{bf}} = t_{\rm{d}}$. 

\begin{figure}
    \centering
    \includegraphics[width=14cm]{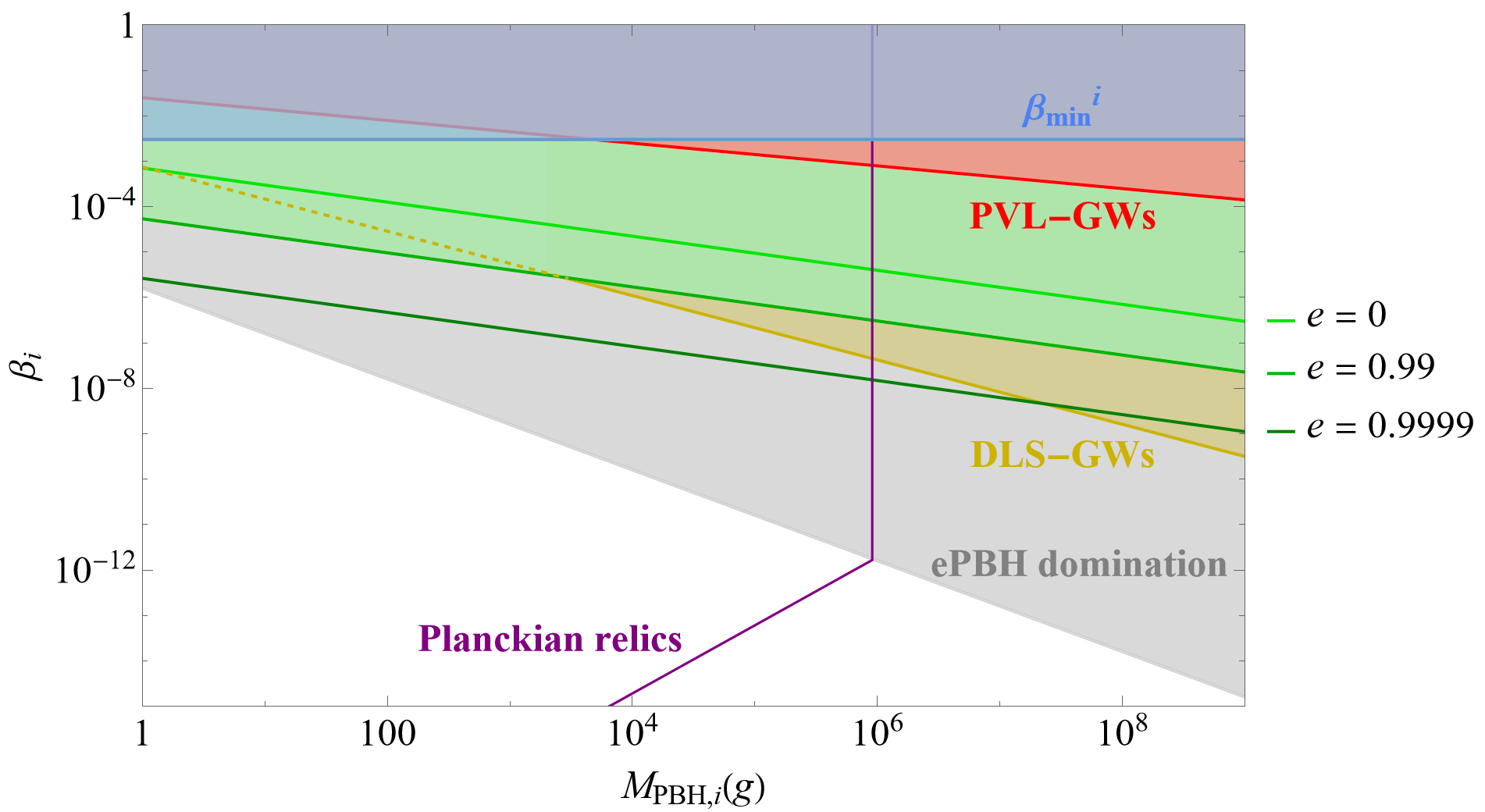}
    \caption[Indirect constraints and limits on the fraction of the universe composed of ultra-light PBHs, denoted as $\beta_{\rm{i}}$.]{\label{figbetamer} Indirect constraints on the fraction of the universe composed of ultra-light PBHs, denoted as $\beta_{\rm{i}}$, as a function of their initial mass. In addition to the information shown in \cref{figGWs}, we include binary merger limits (green) for different eccentricities (above which the PBHs will merge before evaporating) and the minimum allowed value of $\beta_{\rm{min}}^{\rm{i}}$ required for the condition $\Gamma_C = H$ to hold, see \cref{eqBetamin}. These 4 lines do not represent an observational constraint on $\beta_i$ but show where the impact of mergers cannot be neglected when making constraints. The yellow DLS line shows the tentative constraint for GWs from the sudden transition from a matter-dominated to a radiation-dominated era \cite{domenech2021gravitational}.
    The red shaded region shows the backreaction effect from \cite{papanikolaou2021gravitational}, which vanishes when a cutoff is applied to nonlinear density perturbations.}
\end{figure}

The merger time \cite{peters1964gravitational} is
\begin{equation}
\label{eqtmer}
    t_{\rm{mer}} = \frac{3}{85}\frac{a_{x}^{4}}{2 M_{\rm{PBH,i}}^3} (1- e^{2})^{7/2} \left(\frac{c^5}{G^3}\right) \rm{s},
\end{equation}
which depends on the semi-major axis of the newly formed binary \cite{ding2023merger}:
\begin{equation*}
a_x = \alpha \left(\frac{3 M_{\text{PBH,i}}}{4 \pi \rho_{\rm{BG}}}\right)^{1/3} 
\left(\frac{X}{f_{\rm{PBH}}^{\rm eff}}\right)^{4/3},
\end{equation*}
where $\alpha \simeq 0.1$ \cite{franciolini2022assess}, $X \equiv (\xi/\bar{\xi})^3$, and $f_{\rm{PBH}}^{\rm{eff}} = 1$ during the ePBHd era, since PBHs constitute the entire matter content when they dominate, with the superscript ‘eff’ indicating that this definition differs from the standard $\fpbh$. As we are considering the typical case, ($\xi = \bar{\xi}$), the semi-major axis does not depend on $\bar{\xi}$ to good approximation. The eccentricity of the orbit is $e$. We consider eccentricity values ranging from $e = 0$ (a circular orbit) to $e = 0.9999$ (based on the average distribution of the dimensionless angular momentum $j$ for early binaries from the simulation in \cite{delos2024structure}), we also consider $e = 0.99$ as an intermediate case. When $t_{\rm{ev}} > t_{\rm{mer}}$, mergers occur, setting an interesting threshold for PBH abundance in the ultra-light mass regime (see \cref{figbetamer}). Our analysis indicates that most PBHs forming binaries will merge before evaporating, even in the least efficient case of zero eccentricity ($e = 0$), which corresponds to the slowest possible merger.

The impact of the primordial merger limit might broaden the mass function and hence reduces the induced GW constraints from \cite{domenech2021gravitational} on the abundance of ultra-light PBHs. Below the merger limit, these constraints likely hold for $\beta_{\rm{i}}$ within the intermediate range but not for larger values, leading to intriguing and non-trivial behavior.
Primordial mergers also alter the lifetime of the resulting PBHs. Although a naive estimate suggests an eightfold increase in the lifetime, the actual extension is reduced by GW emission and faster evaporation of spinning black holes \cite{taylor1998evaporation, abbott2016observation}. This shift affects PBH constraints as discussed later and decreases the number of Planck-mass relics.

The most related work we are aware of is Hooper et al.~\cite{hooper2020hot}, although it does not address the primordial formation of binaries or its consequences. Instead, it considers binary capture long after formation, assuming PBHs move and capture each other randomly rather than forming binaries immediately. The GW signal calculated in their paper peaks at very high frequencies, as it primarily arises from mergers at small separations, making detection extremely challenging. 

Therefore, a detailed comparison with \cite{hooper2020hot} is necessary to determine the dominant mechanism for binary formation in the early universe.  Following their paper, black holes merge efficiently when their binary capture rate ($\Gamma_C$) exceeds the Hubble rate ($H$), where 
\begin{equation}
\label{eqGammaCH}
    \frac{\Gamma_C}{H} \simeq 
    \frac{135 \, G}{16\, \pi \, \nu^{11/7}}
   \left[ \frac{\beta_{\rm{i}} \, (\frac{t}{t_{\rm{i}}})^{1/2}}{(1+\beta_{\rm{i}} \, (\frac{t}{t_{\rm{i}}})^{1/2})^{1/2}}\right]
   \frac{ M_{\rm{PBH,i}}}{t} \times c^{-10/7},
\end{equation}
where $\nu$ denotes the relative velocity of PBHs, taken as $\nu = 10^{-3}c$ following \cite{hooper2020hot}. To determine the values of $\beta_{\rm{i}}$ 
that satisfy the condition $\Gamma_C = H$, it is useful to note that the ratio $\Gamma_C/H$ decreases with time during radiation domination, hence we need to evaluate it at the initial time. An unrealistic assumption is that binary formation occurs as soon as the black hole pair enter the horizon. 
However, a more realistic choice would be  evaluating it when the gravitational attraction matches the expansion \cref{eqbalance}, 
allowing us to express \cref{eqGammaCH} in terms of \cref{eqtbf}, which shows that  $\Gamma_C = H$ is only possible if
\begin{equation}
    \label{eqBetamin}
    \beta^{\rm{i}}_{\rm{min}} \simeq 3 \times 10^{-3},
\end{equation}
independent of $M_{\rm{PBH,i}}$, as illustrated by the blue line ($\beta^{\rm{i}}_{\rm{min}}$) in \cref{figbetamer}.

For $\beta_{\rm{i}} < \beta^{\rm{i}}_{\rm{min}}$, binaries will not form via this mechanism, making the scenario considered in \cite{hooper2020hot} largely irrelevant, under the assumption that PBHs form in a random, unclustered manner. 
However, for $\beta_{\rm{i}} > \beta^{\rm{i}}_{\rm{min}}$, the dominant binary formation mechanism becomes less clear. Additionally, some PBHs may never merge, while others might undergo multiple mergers, see e.g.~\cite{Holst:2024ubt}.
Although estimating the number of PBHs in each category is highly challenging, the probability of multiple mergers, while small, becomes significant for sufficiently massive PBHs. Their initial mass may allow them to survive until BBN, where constraints are particularly stringent.

To assess the viability of such merger scenarios and identify the parameter space consistent with BBN constraints, one must estimate the properties of the remnant black holes following mergers. This includes accounting for both the mass loss due to GW emission and the spin acquired by the remnant. In our analysis, we adopt a representative spin value of $a_{\rm{s}} = 0.6864$ for the remnant resulting from the merger of two non-spinning black holes of equal mass, as derived in \cite{hofmann2016final,delos2024structure}.
Using the results of \cite{arbey2020evolution}, we estimate the resulting remnant lifetime to be reduced by a factor of 0.76 compared to a black hole with the same mass but zero spin. The mass loss during mergers is estimated based on numerical relativity results. According to \cite{barausse2012mass}, the maximum mass loss reaches up to 9.95\% when the initial black hole spin is maximal ($a_{\rm{s}}=1$), while \cite{abbott2016observation} reports an average loss of approximately 5\% for initial spins below 0.7. As a realistic representative value we take the mass loss per merger to be 5\%. This allows us to approximate the remnant mass as a factor of 1.9 times the initial mass of an individual black hole. 

\begin{figure}[ht]
    \centering
    \begin{minipage}{0.56\textwidth} 
        \centering
        \includegraphics[width=1\linewidth]{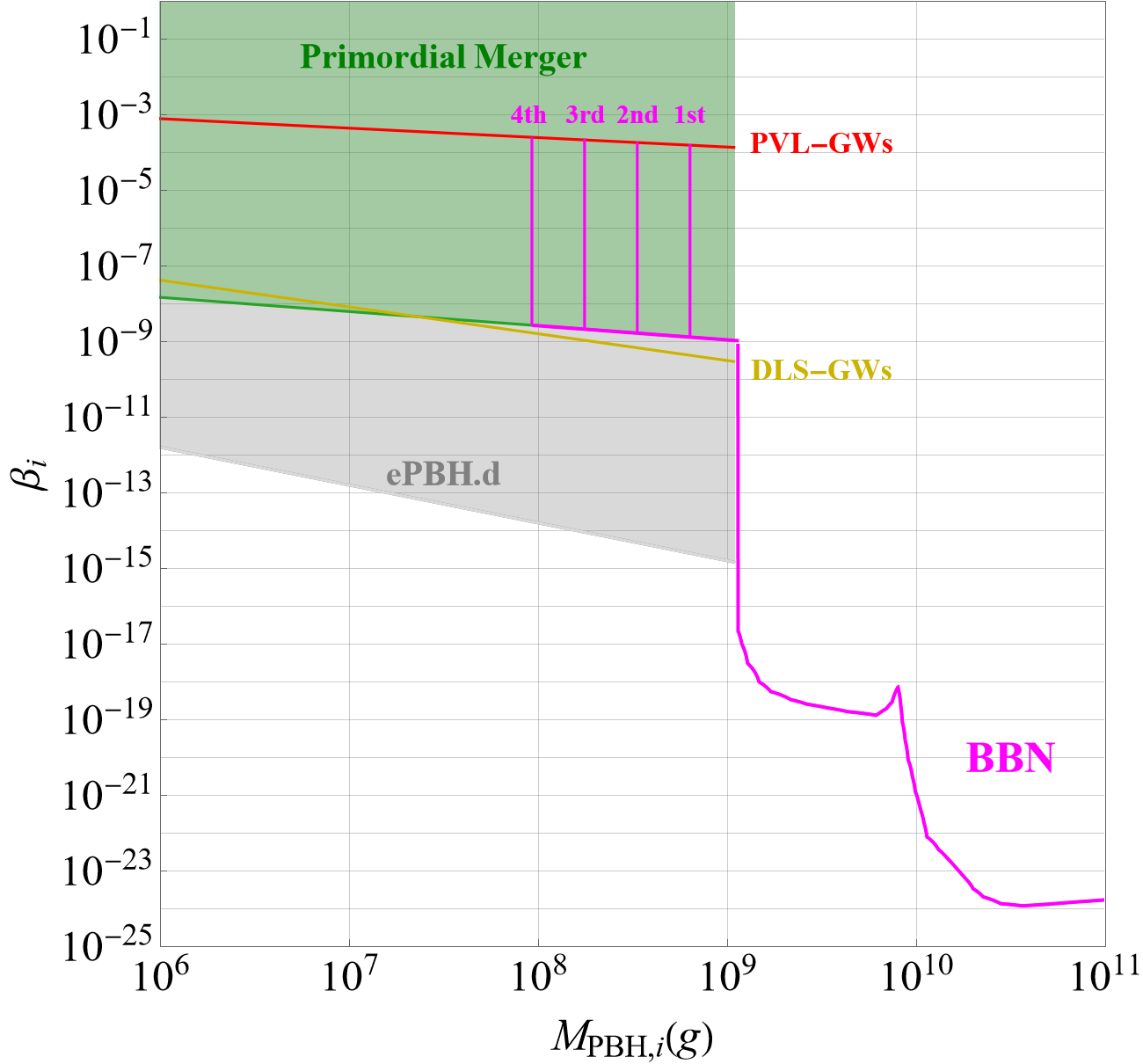}
    \end{minipage}%
    \hspace{1em} 
    \begin{minipage}{0.4\textwidth} 
        \caption[...]{The BBN constraint on the remnants of PBH mergers is shown in magenta, from the first merger (denoted as 1st) up to the fourth (denoted as 4th). For the fourth merger line we assume that at least approximately one in $10^{9}$ PBHs with an initial mass $M_{\rm{PBH,i}} \simeq 10^{8}$ g undergoes four successive mergers, producing remnants that survive long enough to violate BBN bounds. The vertical magenta lines show where the constraints jump by over 4 orders of magnitude to the weaker PVL GW constraint shown in red (or even to $\beta_i=1$ \cite{PhysRevD.105.123023})  for smaller masses which fully evaporate before BBN.}
        \label{fignewBBN}
    \end{minipage}
\end{figure}

By evaluating the initial mass and spin of the post-merger remnant, we can delineate regions of parameter space in which mergers are disallowed due to BBN constraints. Specifically, the lifetime of the remnant is given by $0.76 \times (1.9)^3 t_{\rm{ev}}$ after one merger, and $0.76 \times (1.9)^{3n} t_{\rm{ev}}$ after $n$ successive mergers (assuming each merger is between black holes of the same generation), where $t_{\rm{ev}}$ is the lifetime of the original mass black hole,  \cref{eqtev}. It is important to note that, according to \cite{fishbach2017ligo}, the spin does not significantly increase in subsequent mergers, implying that changes to the lifetime beyond the first merger arise predominantly from cumulative mass growth. The resulting constraints are illustrated in \cref{fignewBBN}, where we show the excluded region of up to a 4th generation black hole based on the post-merger remnant lifetime relative to the BBN bounds. This plot illustrates the number of merger generations required to violate BBN constraints. For example, in the case of a fourth-generation merger, we do not need to assume that most PBHs undergo four mergers. We show only that if even a small fraction, e.g., $10^{-8}$, experiences four mergers, their survival rate could still be sufficient to exceed BBN limits because the BBN limits are so tight. While \cite{Holst:2024ubt} explored scenarios in which a large number of successive mergers significantly constrain the allowed parameter space, we restrict our analysis to a more conservative case of fewer mergers.

We note that a recent paper has claimed that relativistic effects in the early universe lead to ultra-light PBHs accreting significantly before decaying \cite{Das:2025vts}. This also leads to the BBN constraint applying to lower masses than previously expected.

\subsection{A comprehensive constraint plot on the initial abundance of PBHs}
\label{secbetaplot}
By synthesizing insights from various sources, we provide a comprehensive perspective that considers the suppression effect of the ePBHd era, rapid transitions between cosmological epochs, and the formation and mergers of primordial binaries. To ensure that the formation temperature does not exceed the inflationary energy scale, the mass of a PBH must satisfy $M_{\rm{PBH,i}} \gtrsim 1$g, which follows from the constraint on the Hubble parameter during inflation, $H_{\rm{inf}} \lesssim 10^{14}$ GeV \cite{channuie2015strong}. 

\begin{figure}
    \centering
    \includegraphics[width=15.5cm]{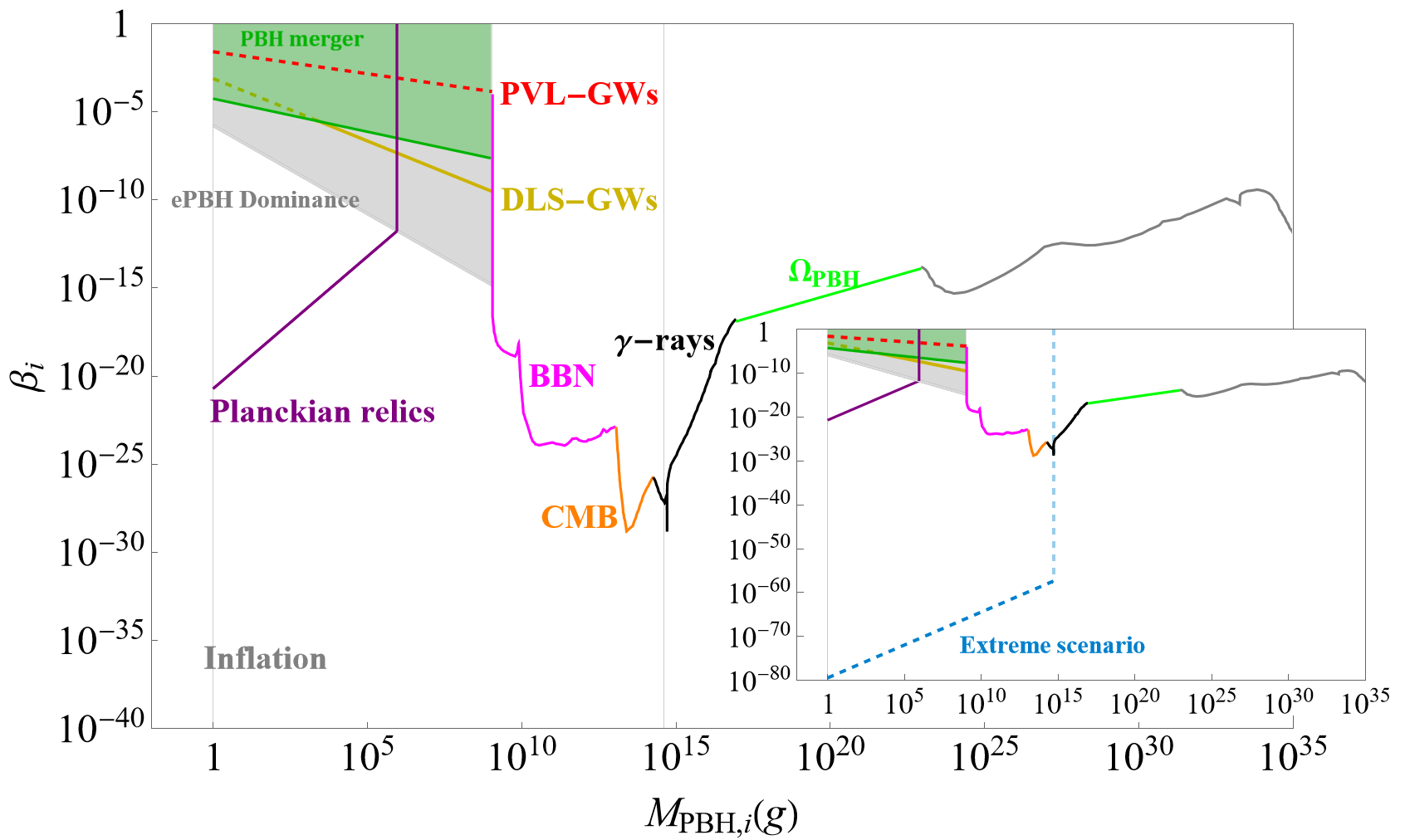}
    \caption[...]{\label{figboth} Constraints on the initial PBH fraction, $\beta_{\rm{i}}$, as a function of initial mass, $M_{\rm{PBH,i}}$ (in grams), considering both evaporated and non-evaporated PBHs. In the main plot, the strongest constraint for $M_{\rm{PBH,i}} \lesssim 10^{6}$ g arises from PBH evaporation into Planck-mass relics for $f_{\rm{PBH}}=1$. Following this, constraints from induced GWs due to PBH evaporation (yellow) 
    and from PBH inhomogeneities (dashed red) are shown, with the latter vanishing when incorporating the suppression factor from \cref{eqR}. These GW constraints depend on the onset of ePBHd, indicated by the shaded region. The green-shaded area represents the merger limit for $e = 0.99$, replacing the left orange dashed GW from transition line. The gray line at $M_{\rm{PBH,i}} \sim 10^{15}$ g marks the maximum mass that could have fully evaporated by today. Constraints from BBN, CMB, $\gamma$-rays, partially and non-evaporated PBHs, adapted from \cite{carr2021constraints}, are displayed on the right. The inset plot illustrates an extreme constraint on $\beta_{\rm{i}}$ under the assumption that no PBHs have evaporated within the observable universe. 
    }
\end{figure}

Building on these considerations, we present \cref{figboth}, which combines direct constraints from \cite{carr2021constraints}, based on observational limits such as gamma-ray backgrounds and the CMB, with indirect constraints from gravitational wave studies, ultra-light PBH mergers, and Planck-mass relic production. The plot shows the initial PBH fraction, $ \beta_{\rm{i}}, $ as a function of the initial PBH mass, $ M_{\rm{PBH,i}}, $ in grams.
Notably, the definition of $\beta_{\rm{i}}$ is simplified compared to \cite{carr2021constraints}, who use $\beta^{'}$, which is related to $\beta_i$ via
\begin{equation*}
    \label{eqbetadash}
    \beta^{'} = 
     \left(\frac{\gamma}{0.2}\right)^{1/2}
    \left(\frac{g_{\ast}}{106.75}\right)^{-1/4}
    \left(\frac{h}{0.67}\right)^{-2} \beta_{\rm{i}},
\end{equation*}
where $\gamma = 0.2$  is a prefactor relating the initial PBH mass to the horizon mass, $g_{\ast}= 106.75$ is the number of relativistic degrees of freedom at PBH formation, and $h = 0.67$ is the dimensionless Hubble constant. 

We also show the extreme constraint of never having had a single PBH evaporate in the observable universe, which can be motivated by the potential destabilization of the Higgs vacuum of the universe, which yields the tightest possible constraint \cite{Cole:2017gle,nakama2018limits,hamaide2024primordial, byrnes2024robust}, as shown in \cref{figboth}. If future Standard Model measurements confirm the electroweak vacuum's metastability—ruling out PBH evaporation—this extreme constraint would entirely exclude scenarios with an early PBH-dominated era or PBH relics.

\section{Constraining the fraction of Dark Matter in Primordial Black Holes   }
\label{secfPBH}
PBHs are a unique dark matter candidate that require no new physics or particles. 
Their present-day abundance is described by the fraction $f_{\rm{PBH}}$ \cite{green2021primordial} (see \cref{eqBetaNonevap,eqbetaD}). If no relics form, PBHs with initial masses below $M_{\rm{PBH,i}} \approx 10^{15}$g fully evaporate and cannot contribute to dark matter, underscoring the relevance of relics for ultra-light PBH scenarios.

\subsection{Tightening the observational constraints using relics  }
\label{sectighterCons}
PBHs can Hawking evaporate into a range of standard and beyond-the-standard model particles. As shown in \cref{figboth}, the strongest constraints on $\beta_{\rm{i}}$ lie in the mass range $10^9\,\mathrm{g} < M_{\rm{PBH,i}} < 10^{15}\,\mathrm{g}$, mainly due to effects on BBN and the CMB. We will show that inclusion of Planck-mass relics result in even tighter $f_{\rm{PBH}}$ constraints. The direct observational constraints on $\beta_{\rm{i}}$ are not influenced by the presence or absence of Planck-mass relics resulting from PBHs. This is because the initial PBH mass ($M_{\rm{PBH,i}}$) is much larger than the final mass after evaporation ($M_{\rm{Pl}}$). The energy released during the evaporation process is given by $E = M_{\rm{PBH,i}}c^2 - M_{\rm{Pl}}c^2$, which remains essentially unchanged. As a result, it is reasonable to conclude that these constraints on $\beta_{\rm{i}}$ can be translated into equivalent constraints on $f_{\rm{PBH}}$. 

To the best of our knowledge, most studies largely overlook the role of relics in the mass range $10^{9}\rm{g} < M_{\rm{PBH,i}} < 10^{15}\rm{g}$. However, constraints on $f_{\rm{PBH}}$ are highly sensitive to this range since, without relics, $f_{\rm{PBH}}=0$ following evaporation. Including relics $f_{\rm{PBH}}$ satisfies 
\begin{equation}
\label{eqFPBHfromBcs}
\qquad\qquad\qquad\qquad\qquad\qquad f_{\rm{PBH}} = \left(\frac{M_{\rm{pl}}}{M_{\rm{PBH,i}}}\right) 
    \left(\frac{a_{\rm{eq}}}{a_{\rm{i}}}\right)
    \beta_{\rm i}, \,\,\,\,\,\,\,\,\,\,\,\,\,\, \,\,\,\,\,\,\,\,\,\,\,\,( M_{\rm{PBH,i}} < 10^{15}\rm{g}).
\end{equation}
Extremely tight constraints on $f_{\rm{PBH}}$ resulting from assuming Planck-mass relics can form are depicted in part (c) of \cref{figfPBHmi}.

\subsection{The constraint from an ePBHd era on $f_{\rm{PBH}}$  }
\label{secmappingONf}
As previously mentioned, if PBHs dominate before decay then $\beta\simeq1$ temporarily. However, unlike on the fraction $\beta$, this scenario introduces a strong constraint on $f_{\rm{PBH}}$ for the mass range $m_{\rm{min}}^{\rm{i}} = M_{\rm{PBH,i}} \gtrsim 1.5 \times 10^6 \rm{g}$, which is the mass where the Planck-mass relics constraints on $\beta_{\rm{i}}$ intersects with the ePBHd limit in \cref{figbetamer}. This restriction determines the fraction of PBH relics contributing to the dark matter abundance as a function of the initial PBH mass, 
\begin{align*}
    f_{\rm{PBH}} &= 2\left(1+ \frac{\Omega_{\rm{b}}}{\Omega_{\rm{DM}}}\right)\left(\frac{M_{\rm{pl}}}{M_{\rm{PBH,i}}}\right) 
    \left(\frac{a_{\rm{d}}}{a_{\rm{i}}}\right)
    \left(\frac{a_{\rm{eq}}}{a_{\rm{ev}}}\right)
    \beta_{\rm{i}}\\
    &= 2\left(1+ \frac{\Omega_{\rm{b}}}{\Omega_{\rm{DM}}}\right) \left(\frac{M_{\rm{pl}}}{M_{\rm{PBH,i}}}\right) 
    \left(\frac{t_{\rm{d}}}{t_{\rm{i}}}\right)^{1/2}
    \left(\frac{t_{\rm{eq}}}{t_{\rm{ev}}}\right)^{1/2}
    \beta_{\rm{i}}
\end{align*}
\begin{equation}
    \label{eqfPBHd}
     \,\,\,\,\,\,\,\,\,\, \,\,\,\,\,\,\,\,\,\,\,\,\,  \,\,\,\,\,\,\,\,\,\,
    f_{\rm{PBH}} \simeq 
    2.8 \times 10^{15} \left(\frac{M_{\rm{PBH,i}}}{\rm{g}}\right)^{-5/2}, \,\,\,\,\,\,\,\,\,\,\,\,\,\,\,\,\,\,\,\, \,\,\,\,\,\,\,\,\,\,  \,\,\,\,\,\,\,\,\,\, \,\,\, (\beta_{\rm{i}} > \beta_{\rm{cri}})
\end{equation}
where $\beta_{\rm{cri}}$ was defined in \cref{eqBibiggerBcri} and corresponds to the threshold for ePBHd.

The equation for $f_{\rm{PBH}}$ now depends solely on $M_{\rm{PBH,i}}$ and is independent of $\beta_{\rm{i}}$. Consequently, even in cases where there are no observational constraints on $\beta$, we can still constrain $f_{\rm{PBH}}$, as shown in part (b) of \cref{figfPBHmi}. The lower limit of $M_{\rm{PBH,i}}$ is determined by the intersection of the ePBHd line (depicted in blue) with the ``relics $f_{\rm{PBH}}=1$" window, which will be elaborated on in \cref{secrelicsWindow}.

\begin{figure}
\centering
\includegraphics[width=10cm]{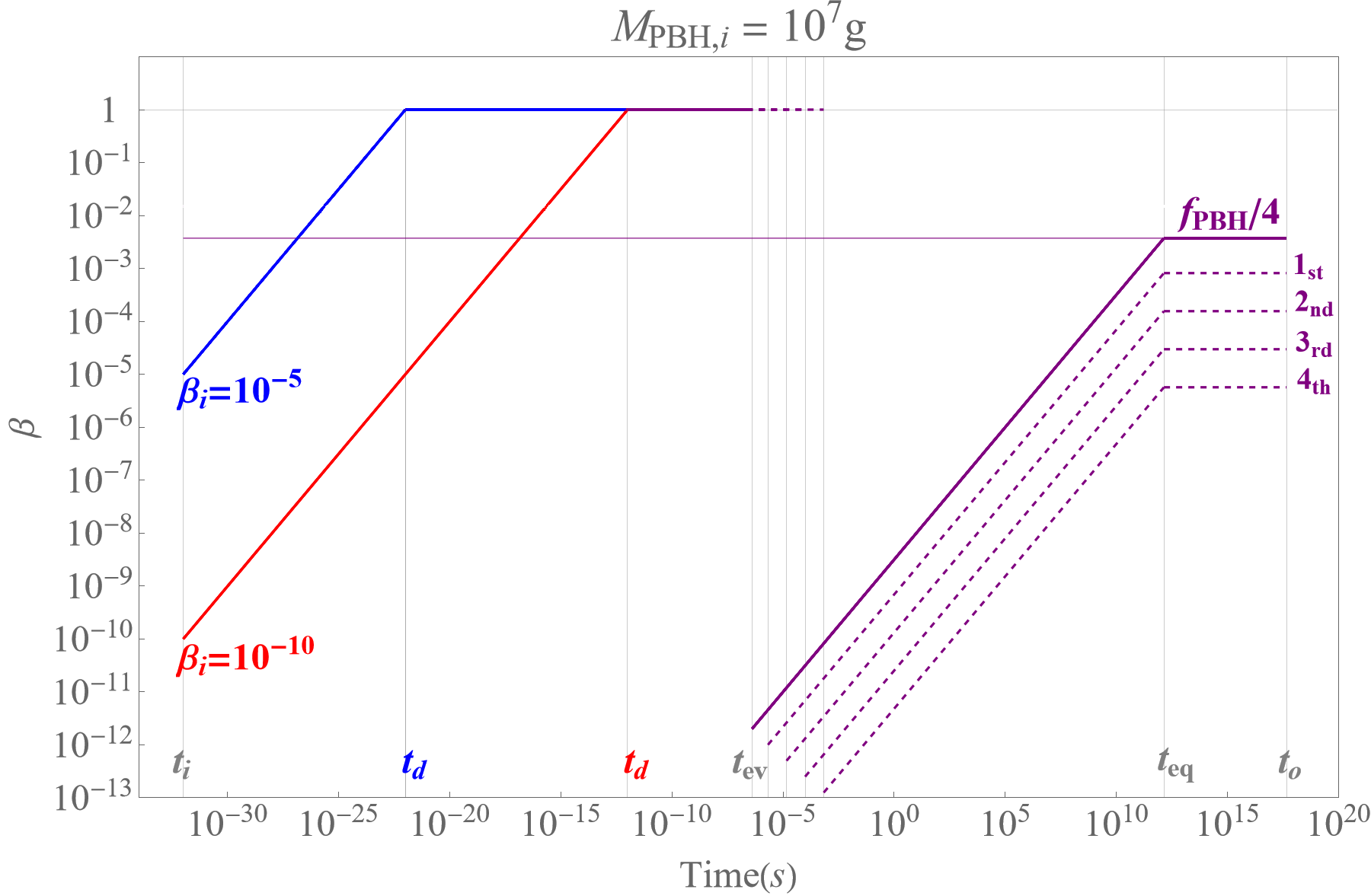}
\caption[...]{\label{figbetatime} The evolution of the PBH fraction, $\beta$, over time for an initial PBH mass of $M_{\rm{PBH,i}} = 10^{7}$g. The plot illustrates the evolution with two different initial abundances: $\beta_{\rm{i}} = 10^{-10}$ (red) and $\beta_{\rm{i}} = 10^{-5}$ (blue). During ePBHd, $\beta=1$ for both cases before the PBHs evaporate, resulting in the same value of $f_{\rm{PBH}}$ today. Note that due to dark matter making up about 0.25 of the total energy density of the universe, then $ fPBH \simeq 4\beta_{eq}$. Additionally, including mergers would change the second half of the plot as there is less radiation domination after decay time for $\beta$ to grow, as well as reducing the number density by a factor of $2^{n}$ (where $n$ is the number of successful mergers between the same generation).  
}
\end{figure}

Another way to interpret \cref{eqfPBHd} is that, regardless of the evolutionary pathway of the universe towards the current value of $f_{\rm{PBH}}$, the outcome remains the same provided $\beta$ reaches one. For a clearer understanding, refer to \cref{figbetatime}. 

\subsection{The Planck--mass window }
\label{secrelicsWindow}
The constraints on $f_{\rm PBH}$ give rise to  ``windows" of initial PBH masses, within a certain range (for example, the asteroid-mass window between about $10^{17}$g and $10^{22}$g), where no direct observational constraints conclusively rule out the possibility of PBHs fully accounting for dark matter. This window has been the subject of extensive debate, with various approaches attempting to impose limits. Some of these efforts involve dropping the assumption of a monochromatic mass function  \cite{carr2022primordial,Gorton:2024cdm}, or reviewing existing observational data and constraints \cite{sasaki2018primordial,carr2023observational}. Less attention has been devoted to the implications of PBH evaporation into relics, particularly in the context of PBHs contributing to dark matter. 

\begin{figure}
\centering
\includegraphics[width=15cm]{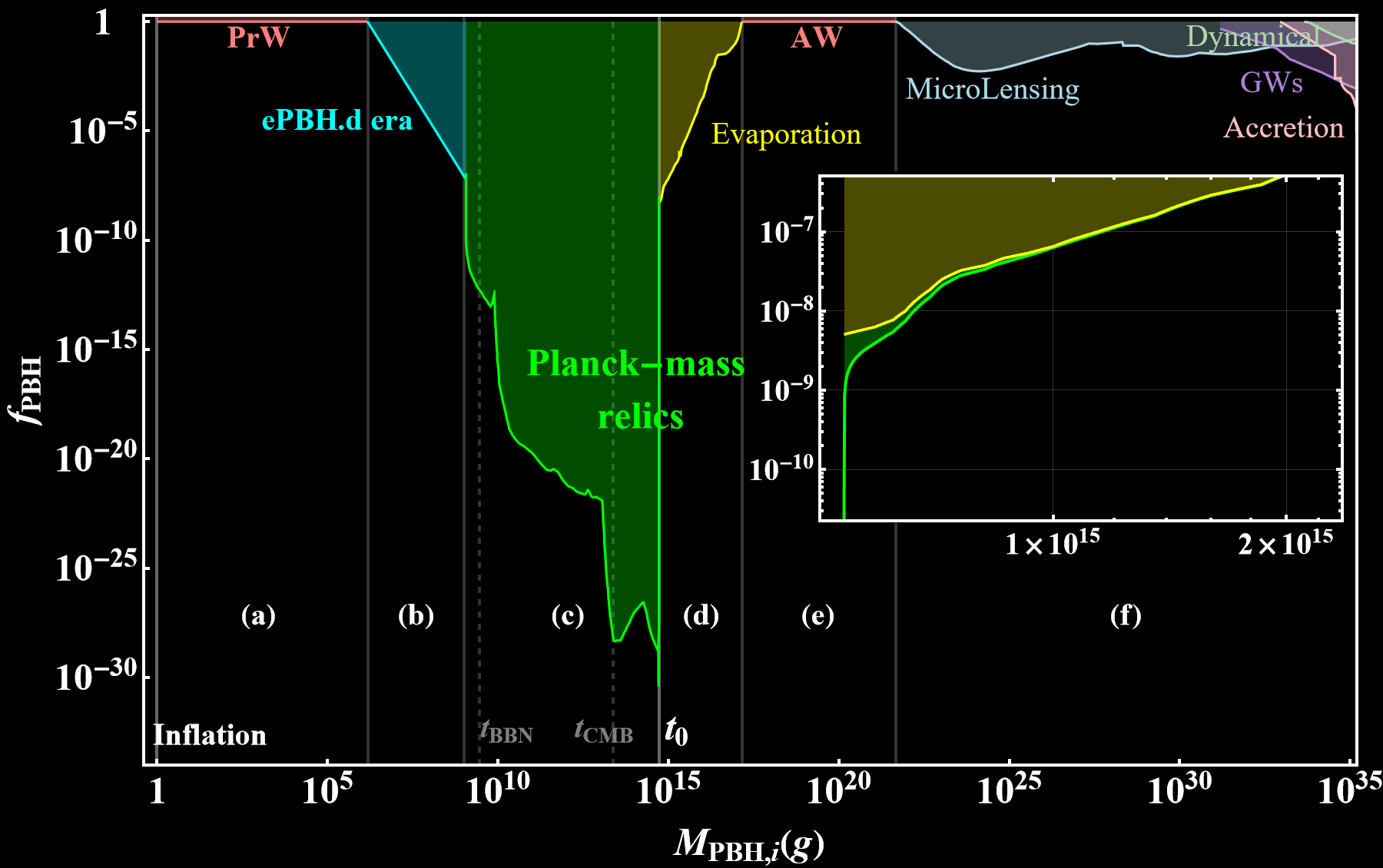}
\caption[...]{\label{figfPBHmi} The constraints on the fraction of dark matter composed of PBHs, $f_{\rm{PBH}}$, as a function of their initial mass, $M_{\rm{PBH,i}}$ (in grams). Moving rightward (with increasing $M_{\rm{PBH,i}}$), the line at $M_{\rm{PBH,i}} = 1$ g corresponds to PBHs forming instantly after inflation ends, assuming the highest possible energy scale of inflation. To the right, part (a) illustrates a window where dark matter can entirely consist of Planck-mass relics (PrW), with the upper limit defined by part (b) (cyan), which represents PBH domination (ePBHd) \cref{eqfPBHd}. Note that even though the constraints on $\beta_i$ are weaker than ePBHd, the resulting constraint on $f_{\rm{PBH}}$ does not change. Further right, part (c) (green) shows constraints derived by mapping $\beta$ into $f_{\rm{PBH}}$, assuming PBHs evaporate into Planck-mass relics - see \cref{eqFPBHfromBcs}. Parts (d), (e), and (f), adapted from \cite{green2021primordial}, show additional constraints: (d) describes evaporating PBHs whose mass has not significantly decreased since formation, (e) highlights an asteroid-mass window (AW) where non-evaporating PBHs can account for all dark matter, and (f) presents constraints on stable PBHs. If PBHs do not leave relics after evaporation, $f_{\rm{PBH}} = 0$ for $M_{\rm{PBH,i}}$ values to the left of the vertical line at $t_0$, as these PBHs would have fully evaporated.
}
\end{figure}

By taking the evaporation into relics seriously \cite{macgibbon1987can,Carr1994,carr2016primordial}, we see an additional window for relics with $f_{\rm{PBH}} = 1$, 
while simultaneously tightening the observational constraints on PBHs in the mass range of $10^{9}$g $\lesssim M_{\rm{PBH,i}} \lesssim 10^{15}$g, particularly around the time of evaporation (see \cref{secmappingONf}). This is encapsulated in our first-of-its-kind master plot, shown in \cref{figfPBHmi}, which summarizes the most stringent constraints within the range of $1$g $< M_{\rm{PBH,i}} < 10^{35}$g, incorporating the assumption of Planck-mass relics.

A discontinuity of approximately 20 orders of magnitude can be observed in \cref{figfPBHmi} at the mass scale which evaporates today ($t_0$), which stems from assuming the relic mass to be the Planck mass. In reality this apparent discontinuity is smooth  because PBHs with masses only slightly above the minimum value $M_*$ which evaporate to a relic today will have lost an appreciable fraction of their initial mass by today. However, the discontinuity is sharp because only a small range of initial masses lead to $M_{\rm pl}\ll M_0\ll M_i$ where $M_0$ is today's mass. To see why, consider one with initial mass $M_i=\Psi M_*$ for $\Psi\gtrsim1$. Their mass has reduced to $M_0=\sqrt[3]{M_i^3-M_{*}^3}=M_*\sqrt[3]{\Psi^3-1}$ so e.g.~$\Psi=2$ corresponds to a PBH which has lost only 4\% of its initial mass. Taking this into account lead to only a modest shift in the evaporation constraint which is illustrated in the zoomed-in panel of \cref{figfPBHmi}. 

If the fraction $f_{\rm{PBH}}$ reaches $1$, it implies that the entirety of the current dark matter density is composed of PBHs (see part (a)). By calculating the minimum initial PBH mass required for domination, we identify a constraints-free ``Planck-mass relics window" where $f_{\rm{PBH}} = 1$. The width of this window depends on the inflation model's capacity to generate such small initial PBH masses: a lower energy scale at the end of inflation delays reheating, resulting in a larger horizon mass and larger minimum PBH mass.

This relic window (corresponding to initially light PBHs) is arguably more naturally reached than the non-evaporated PBH window (e.g., the asteroid-mass window), as the power spectrum is constrained to be very small on large scales. While models often assume a peaked power spectrum which may require fine-tuning \cite{Cole:2023wyx}, (but see also \cite{Stamou:2023vft}), a more natural scenario might involve a spectrum that grows continuously, reaching its maximum value near the end of inflation, corresponding to the smallest scale, see e.g.~\cite{Carr:2017edp,Briaud:2023eae,Wilkins:2023asp}. Such a scenario would naturally favor the formation of very light PBHs. If the amplitude of the power spectrum becomes sufficiently high, the resulting abundance of low-mass PBHs could arguably more easily lead to $\beta_i \sim 1$. 
If the spectrum grows steadily without decaying, the smallest scales—and consequently, the lightest PBH masses—emerge as the most significant contributors to dark matter.

\section{Conclusion}

Primordial black holes (PBHs), a hypothesised class of black holes formed via the collapse of enhanced inflationary perturbations, offer a unique probe into early-Universe physics. PBHs with initial masses below $\sim10^9\,\mathrm{g}$ evaporate prior to the onset of Big Bang Nucleosynthesis (BBN), allowing them to evade most observational constraints. Consequently, this mass range remains one of the least explored regions of the PBH parameter space.

In this study, we revisit the pre-BBN regime and provide new constraints on the initial PBH mass fraction, $\beta_i$, and the present-day PBH abundance, $f_{\rm PBH}$, within this weakly constrained window. Our analysis incorporates the effects of Planck-scale relics and a possible early PBH-dominated phase. We find that PBHs forming relics between BBN and the present day are subject to significantly tighter constraints—by up to 25 orders of magnitude—compared to stable-mass PBHs, as illustrated in \cref{figfPBHmi}. Notably, there exists a viable parameter region in which Planck relics could constitute the entirety of dark matter, specifically for initial masses $M_{\rm PBH,i} \lesssim 10^6\,\mathrm{g}$. A comprehensive set of constraints on the formation fraction $\beta_i$ is presented in \cref{figboth}.

We further analyse the impact of early PBH binary formation on gravitational wave (GW) signatures and BBN constraints. Our results indicate that early-time mergers—occurring prior to Hawking evaporation—can extend BBN limits to lighter PBH masses. In particular, we demonstrate that ultra-light PBHs undergo mergers before evaporation across a broad parameter space, producing longer-lived remnants that are more strongly constrained by BBN. The extent of this constraint enhancement depends on the number of mergers prior to evaporation, as shown in \cref{fignewBBN}. Our analysis shows that early binary formation dominates over late-time capture mechanisms, particularly for small initial mass fractions ($\beta_i \lesssim 10^{-3}$), rendering the latter ineffective in such scenarios, as discussed in \cite{hooper2020hot}. PBH mergers may also broaden an initially monochromatic PBH mass spectrum, thereby suppressing the stochastic GW background expected from a rapid transition from matter to radiation domination.

Our findings suggest that, for PBH masses $M_{\rm PBH} \lesssim 10^9\,\mathrm{g}$, current observational constraints on the formation fraction $\beta_i$ are either weak or absent—potentially even under the simplifying assumption of a monochromatic mass function. This effectively reopens a large region of parameter space, potentially allowing the probing of inflationary scales corresponding to these early-forming PBHs\footnote{We note that \cite{Kim:2025kgu} claim tight constraints via CMB isocurvature bounds provided PBHs do not dominate before decay, but in our view these large scale isocurvature perturbations are only generated if there is at least some primordial non-gaussianity \cite{Tada:2015noa,Young:2015kda,vanLaak:2023ppj}.}. 

Meanwhile, possible mechanisms through which an early PBH-dominated era could lead to the formation of more massive primordial black holes have been explored.  In particular, \cite{Kim:2024gqp} investigates PBH formation at horizon crossing during this epoch, arguing that reduced pressure massively enhances collapse. Their analysis builds upon the pressureless approximation of \cite{Khlopov:1980mg,Harada:2016mhb}, though in practice, the critical overdensity for collapse increases near the end of the matter-dominated phase (see e.g.~\cite{cole2018extreme}). Alternatively, \cite{Holst:2024ubt} explores whether successive mergers during the PBH-dominated phase could generate more massive PBHs, though this scenario requires further investigation via numerical simulations to assess the merger efficiency.

\begin{center}
    {\bf Acknowledgements} 
\end{center}

We would like to thank all those who contributed to this work. In particular, we are grateful to Lucien Heurtier, Robyn Munoz, Xavier Pritchard, Graham White, and Sam Young for their valuable help and discussions. Amirah is supported by the Government of Saudi Arabia through Taibah institution. CB is supported by STFC grants ST/X001040/1 and ST/X000796/1. 





\printbibliography

\end{document}